\documentclass[preprint]{elsarticle}

\usepackage{hyperref}
\usepackage{microtype}
\usepackage{tabularx}

\journal{XXXX}

\begin{document}

\begin{frontmatter}

\title{A Survey on Workflow Satisfiability, Resiliency, and Related Problems}
\author[fbk]{Daniel Ricardo dos Santos\corref{cor}}
\cortext[cor]{Corresponding author}
\ead{danielricardo.santos@gmail.com}
\author[fbk]{Silvio Ranise}

\address[fbk]{
Security \& Trust \\
Fondazione Bruno Kessler \\
Via Sommarive, 18 - Trento, Italy
}

\begin{abstract}
Workflows specify collections of tasks that must be executed under the
responsibility or supervision of human users. Workflow management systems and
workflow-driven applications need to enforce security policies in the form of
access control, specifying which users can execute which tasks, and
authorization constraints, such as Separation of Duty, further restricting
the execution of tasks at run-time. Enforcing these policies is crucial to
avoid frauds and malicious use, but it may lead to situations where a
workflow instance cannot be completed without the violation of the policy.
The Workflow Satisfiability Problem (WSP) asks whether there exists an
assignment of users to tasks in a workflow such that every task is executed
and the policy is not violated. The WSP is inherently hard, but solutions to
this problem have a practical application in reconciling business compliance
and business continuity. Solutions to related problems, such as workflow
resiliency (i.e., whether a workflow instance is still satisfiable even in
the absence of users), are important to help in policy design. Several
variations of the WSP and similar problems have been defined in the
literature and there are many solution methods available. In this paper, we
survey the work done on these problems in the past 20 years.
\end{abstract}


\end{frontmatter}

\section{Introduction}
\label{sec:intro}

A workflow specifies a collection of tasks, whose execution is initiated by
humans or software agents executing on their behalf, and the constraints on the
order of execution of those tasks. Workflows represent a repeatable and
structured collection of tasks designed to achieve a desired goal. They are
used to model business processes that capture the activities that must be
performed in a business setting to provide a service or product.

Workflow enactment services provide the run-time environment that controls and
executes workflows. Both functional and non-functional requirements need to be
considered when implementing workflow-based applications~\cite{aalst2013}.
Workflow patterns for control-flow, data-flow, and organizational
resources~\cite{aalst2003,russell2005,russell2005er} can be used to
elicit functional requirements, whereas security-related dependencies are
specified in workflows as authorization policies and additional constraints on
the execution of the various tasks.

Authorization policies specify that, in an organization, a workflow task is
executed by a user who should be entitled to do so; e.g., the teller of a bank
may create a loan request, whereas only a manager may accept it. Additional
authorization constraints are usually imposed on task execution, such as
Separation or Binding of Duties (SoD or BoD) whereby two distinct users or the
same user, respectively, must execute two tasks. Workflows equipped with an
authorization policy and constraints may be called
``security-sensitive''~\cite{armando2009}.

While the enforcement of authorization policies and constraints is fundamental
for security~\cite{leitner2014}, it may also lead to situations where a
workflow instance cannot be completed because no task can be executed without
violating either the authorization policy or the constraints. These deadlock
situations emphasize the conflict between business compliance and business
continuity. Business compliance states that the business processes must follow
the modeled workflows, respecting control-flow constraints, authorization
policies and authorization constraints. Business continuity, on the other hand,
states that the business must not stop even in adverse conditions, e.g., in the
absence of authorized users. These conflicts may be resolved by an
administrator granting additional permissions to a user, which violates the
original intended policy and therefore hurts compliance. In alternative, an
administrator can cancel the execution of a business process instance, which
violates business continuity. An ideal solution is to avoid that any instance
execution ever reaches a situation where this choice must be made.

The \emph{Workflow Satisfiability Problem} (WSP) consists of checking if there
exists an assignment of users to tasks such that a security-sensitive workflow
successfully terminates while satisfying all authorization constraints. The
run-time version of the WSP consists of answering sequences of user requests at
execution time and ensuring successful termination together with the
satisfaction of authorization policies and constraints. Such problems have been
studied in several papers and there are many available solutions. Other related
problems have also been studied in the literature. \emph{Workflow Resiliency}
amounts to checking if a workflow can still be satisfied even in the absence of
a certain number of users, while \emph{Workflow Feasibility} concerns the
existence of a possible configuration of the authorization policy (considering,
e.g., delegation or administrative policies) in which the workflow is
satisfiable.

In this paper, we present a survey of the most relevant work in the areas of
workflow satisfiability, resiliency, and related problems. Despite the
number of published papers in these areas (Google Scholar returned 136 results
to the query ``workflow satisfiability'' on June, 2017) and the considerable
time span of the research (20 years, considering the work of Bertino et
al. in 1997~\cite{bertino1997} as the first), to the best of our knowledge,
there is only one previous survey on WSP~\cite{holderer2015}. That survey,
however, is rather brief and only considers papers published until May 2014
(out of the 136 results described above, 59 have been published after 2015).
There is also a related survey by Leitner and Rinderle-Ma~\cite{leitner2014}
that explores the literature on security in process-aware information systems.
Although there is some overlap between~\cite{leitner2014} and this paper, the
former is very broad, considering several aspects of security. It is also brief
when discussing workflow satisfiability or resiliency (the problems are only
mentioned, there is no discussion about the solutions).

The rest of this paper is organized as follows. Section~\ref{sec:method}
describes the methodology adopted for the survey and some statistics about the
published literature. Section~\ref{sec:spec} describes the specification of
security-sensitive workflows, which is necessary to understand the many
variants of the WSP; Section~\ref{sec:wsp} discusses solutions to the WSP;
Section~\ref{sec:res} does the same for the workflow resiliency problem;
Section~\ref{sec:other} presents other related problems that have been less
studied; and Section~\ref{sec:conclusion} concludes the paper.

\section{Methodology}
\label{sec:method}

The methodology used in this paper is similar to that
of~\cite{leitner2014,holderer2015}, which, in turn, follows widely accepted
guidelines for research synthesis~\cite{cooper2009} and systematic literature
review~\cite{kitchenham2004,kitchenham2007}. The methodology is composed of 4
steps. First, we define the research questions
(Section~\ref{sec:identification}). Second, we perform an extensive literature
search (Section~\ref{sec:search}). Third, we select the relevant papers found
in the previous step (Section~\ref{sec:selection}). Fourth, we classify the
literature in terms of problems and solutions (Section~\ref{sec:results}).

\subsection{Research identification}
\label{sec:identification}

Our goal is to identify, classify, and evaluate the literature related to
workflow satisfiability. To do so, we started by examining classic papers in
this domain (e.g.,~\cite{crampton2005,wang2010}) and identified the following
research questions (\textbf{RQ}) to be answered:
\begin{description}
\item[RQ1] What kinds of control-flow patterns and authorization constraints
are currently supported by WSP solutions?
\item[RQ2] What related problems have been defined in the literature and how
are they solved?
\item[RQ3] What are the current research challenges in the WSP and related
problems?
\end{description}

The first question (\textbf{RQ1}) concerns the current state of solutions to
the WSP and its many variants (e.g., ordered WSP~\cite{crampton2013}, valued
WSP~\cite{crampton2017jcs}, and run-time WSP~\cite{bertolissi2015}). An answer
to this question helps us in identifying a broad set of relevant related work.
The second question (\textbf{RQ2}) concerns currently identified problems that
have a strong connection to the WSP (e.g., resiliency~\cite{wang2010},
feasibility~\cite{khan2012}, and minimum users~\cite{roy2015}). An answer to
this question enlarges the set of relevant work by considering other problems.
The third question (\textbf{RQ3}) concerns the identification of challenges and
gaps in the current solutions and points to possible future research
directions.

\subsection{Literature search}
\label{sec:search}

\begin{figure}
\centering
\includegraphics[width=\textwidth]{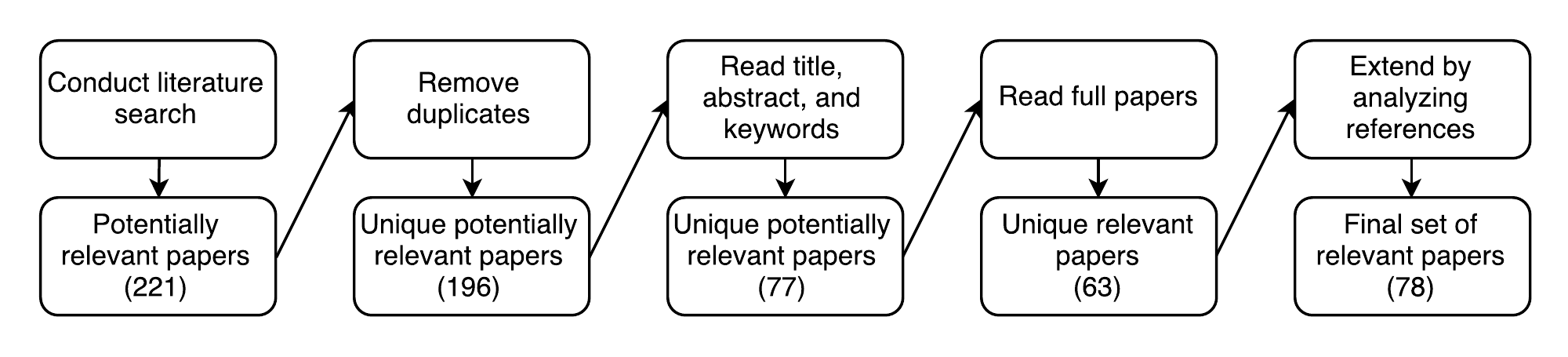}
\caption{\label{fig:steps} Literature search and selection}
\end{figure}

After defining the research questions, we performed the literature search and
selection process, which is shown in Figure~\ref{fig:steps} and detailed in the
next sections. The process was composed of 5 steps (the rectangles at the top
of Figure~\ref{fig:steps}), each producing as output a set of papers (the
rectangles at the bottom of the same Figure). The number of papers in each set
in shown in parentheses and the arrows represent the flow of the steps.
%%%%%%%%%%%%%%%%%%%%%%%%%%%%%%%%%%%%%%%%%%%%%%%%%%%%%%%%%%%%%%%%%%%%%%%%%%%%%%%%
%%%%%%%%%%%%%%%%%%%%%%%%%%%%%%%%%%%%%%%%%%%%%%%%%%%%%%%%%%%%%%%%%%%%%%%%%%%%%%%%

We performed a semi-automated search for literature published until June,
2017 using Google Scholar\footnote{\url{http://scholar.google.com/}}, which
indexes popular computer science research repositories, e.g.,
ACM DL\footnote{\url{http://dl.acm.org/}}, IEEE
Xplore\footnote{\url{http://ieeexplore.ieee.org/}},
SpringerLink{\footnote{\url{http://link.springer.com/}}},
DBLP\footnote{\url{http://dblp.uni-trier.de/}}, and
arXiv\footnote{\url{http://arxiv.org}}. Since Google scholar does not have a
publicly available search API, we used the
\texttt{scholar.py}\footnote{\url{https://github.com/ckreibich/scholar.py}}
crawler. The search terms used were ``workflow satisfiability'', ``workflow
resilience'', ``workflow resiliency'', and ``workflow feasibility''. We
excluded patents from the search. The number of potentially relevant papers
after this first step was 221 (cf. Fig~\ref{fig:steps}). We then automatically
removed duplicate results based on exact title matching (e.g., ``Satisfiability
and Resiliency in Workflow Authorization Systems''~\cite{wang2010} is a result
for the queries ``workflow satisfiability'' and ``workflow resiliency''). The
resulting number of unique potentially relevant papers was 196 (cf.
Fig~\ref{fig:steps}). The next steps in the literature selection were performed
manually.

\subsection{Literature selection}
\label{sec:selection}

We started by reading the title, abstract, and keywords of the results from the
previous step. Many results contain exactly the terms used in the search but
are out of scope for this paper. For instance, terms such as ``feasibility''
and ``resiliency'' have different meanings outside the access control and
security literature. Most PhD theses and papers from the arXiv repository were
also removed, because their contents are described in other publications. After
this step, we were left with 77 papers to read. We then proceeded to fully read
these papers and check if they were relevant to the survey. In some cases, even
though the title and abstract indicate that the paper is relevant, it turns out
not to be. After reading all the papers, we identified 63 unique relevant
papers. Since there are many different ``names'' for the WSP and related
problems (e.g., policy-based deadlock~\cite{kohler2008},
obstruction~\cite{basin2014}), our search did not capture every relevant paper
(we did not include keywords such as ``obstruction'' in our search since it
greatly increases the number of irrelevant results). We then extended the set
of relevant papers by manually checking the references of the papers in the set
and adding the relevant papers that we missed during the initial search (this
process is known in literature review as backward
snowballing~\cite{wohlin2014}). In the end, the total number relevant papers
was 78. We proceeded by analyzing the relevant papers and classifying them.

\subsection{Results}
\label{sec:results}

Table~\ref{tab:papers} shows how we classified the relevant papers.
%%%%%%%%%%%%%%%%%%%%%%%%%%%%%%%%%%%%%%%%%%%%%%%%%%%%%%%%%%%%%%%%%%%%%%%%%%%%%%%%
\begin{table}
\centering
\caption{\label{tab:papers} Classification of the papers}
\begin{small}
\begin{tabular}{|l|l|l|} \hline
Problem & Approach & Papers \\

\hline\hline

WSP & Initial works & \cite{bertino1997} \cite{bertino1999} \cite{tan2004}
\cite{crampton2005} \cite{solworth2006} \cite{wang2007} \cite{wang2010}
\\\cline{2-3}
& FPT algorithms &
\parbox[t]{6cm}{
\cite{crampton2012} \cite{crampton2013tissec} \cite{crampton2013}
\cite{crampton2013faw} \cite{cohen2014faw} \cite{karapetyan2015}
\cite{cohen2014} \cite{cohen2016} \cite{gutin2016} \cite{gutin2016ipl}
\cite{crampton2015ipec} \cite{crampton2016} \cite{crampton2017}
}
\\\cline{2-3}
& Model checking &
\parbox[t]{6cm}{
\cite{xiangpeng2006} \cite{wolter2009} \cite{huth2011} \cite{crampton2012bpmw}
\cite{crampton2012paar} \cite{crampton2014} \cite{bertolissi2013}
\cite{bertolissi2013icitst} \cite{bertolissi2014} \cite{bertolissi2015}
\cite{sacmat2016} \cite{tacas2016} \cite{codaspy2017}
}
\\\cline{2-3}
& Others &
\parbox[t]{6cm}{
\cite{basin2012-sacmat} \cite{burri2012} \cite{basin2012} \cite{basin2011}
\cite{basin2014} \cite{schefer2012} \cite{yang2013} \cite{bo2016}
\cite{hummer2013} \cite{ayed2008} \cite{ayed2009} \cite{jemel2015}
\cite{warner2006} \cite{holderer2016} \cite{combi2016}
}
\\\hline

\hline\hline

Resiliency & Static, dynamic & \cite{wang2010} \cite{jcs2017} \\\cline{2-3}
& Quantitative & \cite{mace2014} \cite{mace2015} \cite{mace2015modelling}
\cite{mace2015impact} \cite{mace2016} \\\cline{2-3}
& Others & \cite{paci2008} \cite{lowalekar2009} \cite{lu2014} \\\hline

\hline\hline

Others & Workflow feasibility & \cite{khan2012} \cite{mehregan2016}
\\\cline{2-3}
& WSP with delegation & \cite{crampton2008} \cite{crampton2008sac}
\cite{crampton2010} \cite{el2012} \cite{el2013} \\\cline{2-3}
& Minimum users & \cite{roy2015} \cite{kohler2008} \cite{dbsec2015}
\cite{jcs2017} \cite{crampton2014} \\\cline{2-3}
& Purpose & \cite{jafari2014} \cite{demasellis2015} \\\cline{2-3}
& Policy properties & \cite{sun2011} \cite{ranise2014} \cite{garrison2014}
\cite{calzavara2016} \cite{berge2016}
%%DECIDE about the ones below...
\cite{crampton2015}
\cite{crampton2017jcs} \cite{wolter2008} \cite{crampton2016sacmat}
\\\hline
\end{tabular}
\end{small}
\end{table}
%%%%%%%%%%%%%%%%%%%%%%%%%%%%%%%%%%%%%%%%%%%%%%%%%%%%%%%%%%%%%%%%%%%%%%%%%%%%%%%%
There are three main categories of works, based on the problem solved (column
`Problem'): WSP, resiliency, and others. Each category has sub-categories,
based on the approach taken (column `Approach'), e.g., fixed-parameter
tractable (FPT) algorithms and model checking. We list the papers in each
category and sub-category (column `Papers'). Notice that some papers
%%(namely, \cite{wang2010}, \cite{crampton2014} and \cite{jcs2017})
appear in two categories because they solve two problems (e.g.,~\cite{wang2010}
is classified as ``initial works'' for the WSP and as ``static, dynamic'' for
workflow resiliency.)

We describe all the papers in Sections \ref{sec:wsp}, \ref{sec:res}, and
\ref{sec:other}, ignoring formal details, but presenting the main results and
key intuitions behind each work. Before presenting the papers in more detail,
though, we outline the literature in terms of year and venue of publication, as
well as author affiliation.

\begin{figure}
\centering
\includegraphics[width=\textwidth]{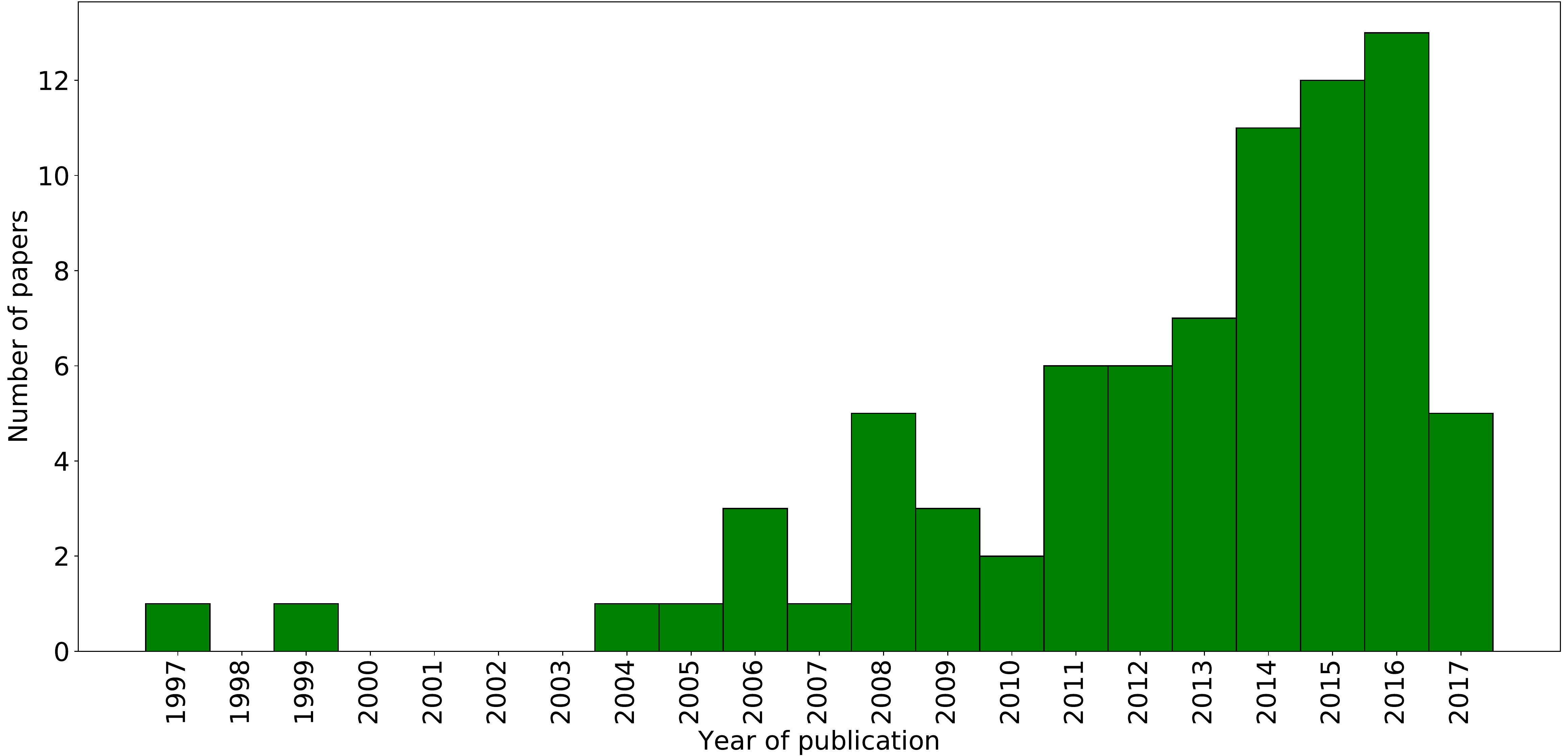}
\caption{\label{fig:papers} Distribution of the papers by year of publication}
\end{figure}

Figure~\ref{fig:papers} shows a histogram with the number of papers by year,
from 1997 to 2017, while Table~\ref{tab:venues} shows the venues and journals
where research in this area is usually published (the Table only shows those
venues and journals with more than one publication).
%%%%%%%%%%%%%%%%%%%%%%%%%%%%%%%%%%%%%%%%%%%%%%%%%%%%%%%%%%%%%%%%%%%%%%%%%%%%%%%%
%%%%%%%%%%%%%%%%%%%%%%%%%%%%%%%%%%%%%%%%%%%%%%%%%%%%%%%%%%%%%%%%%%%%%%%%%%%%%%%%
%%%%%%%%%%%%%%%%%%%%%%%%%%%%%%%%%%%%%%%%%%%%%%%%%%%%%%%%%%%%%%%%%%%%%%%%%%%%%%%%
\begin{table}
\centering
\caption{\label{tab:venues} Journals and venues with more than one publication}
\begin{small}
\begin{tabularx}{\textwidth}{|X|l|l|}
\hline
Venue & Acronym & \# \\\hline\hline
ACM Symposium on Access Control Models and Technologies & SACMAT & 13 \\\hline
ACM Conference on Data and Applications Security and Privacy & CODASPY & 3
\\\hline
Business Process Management Workshops & BPMW & 3 \\\hline
Computer Security Foundations Symposium & CSF & 3 \\\hline
European Symposium on Research in Computer Security & ESORICS & 3 \\\hline
International Frontiers of Algorithmics Workshop & FAW & 3 \\\hline
International Workshop on Security and Trust Management & STM & 3 \\\hline
ACM Symposium on Information, Computer and Communications Security & ASIACCS &
2 \\\hline
IFIP Conference on Data and Applications Security and Privacy & DBSec &
2 \\\hline
International Workshop on Software Engineering for Resilient Systems & SERENE &
2 \\\hline
\hline
Journal & Acronym & \# \\\hline\hline
ACM Transactions on Privacy and Security & TOPS & 6 \\\hline
Journal of Computer Security & JCS & 3 \\\hline
\end{tabularx}
\end{small}
\end{table}
%%%%%%%%%%%%%%%%%%%%%%%%%%%%%%%%%%%%%%%%%%%%%%%%%%%%%%%%%%%%%%%%%%%%%%%%%%%%%%%%
It is clear that this area of research is more popular than ever, with the
number of papers growing each year and 2016 being the most popular year so far
(16 papers). The most popular venue for research is SACMAT, with 13 papers, and
the most popular journal is ACM TOPS, with 6\footnote{SACMAT was called ACM
  Workshop on Role-Based Access Control (RBAC) until 2000, whereas TOPS was
  called ACM Transactions on Information and System Security (TISSEC) until
  2017}.

\begin{figure}
\centering
\includegraphics[width=.3\textwidth]{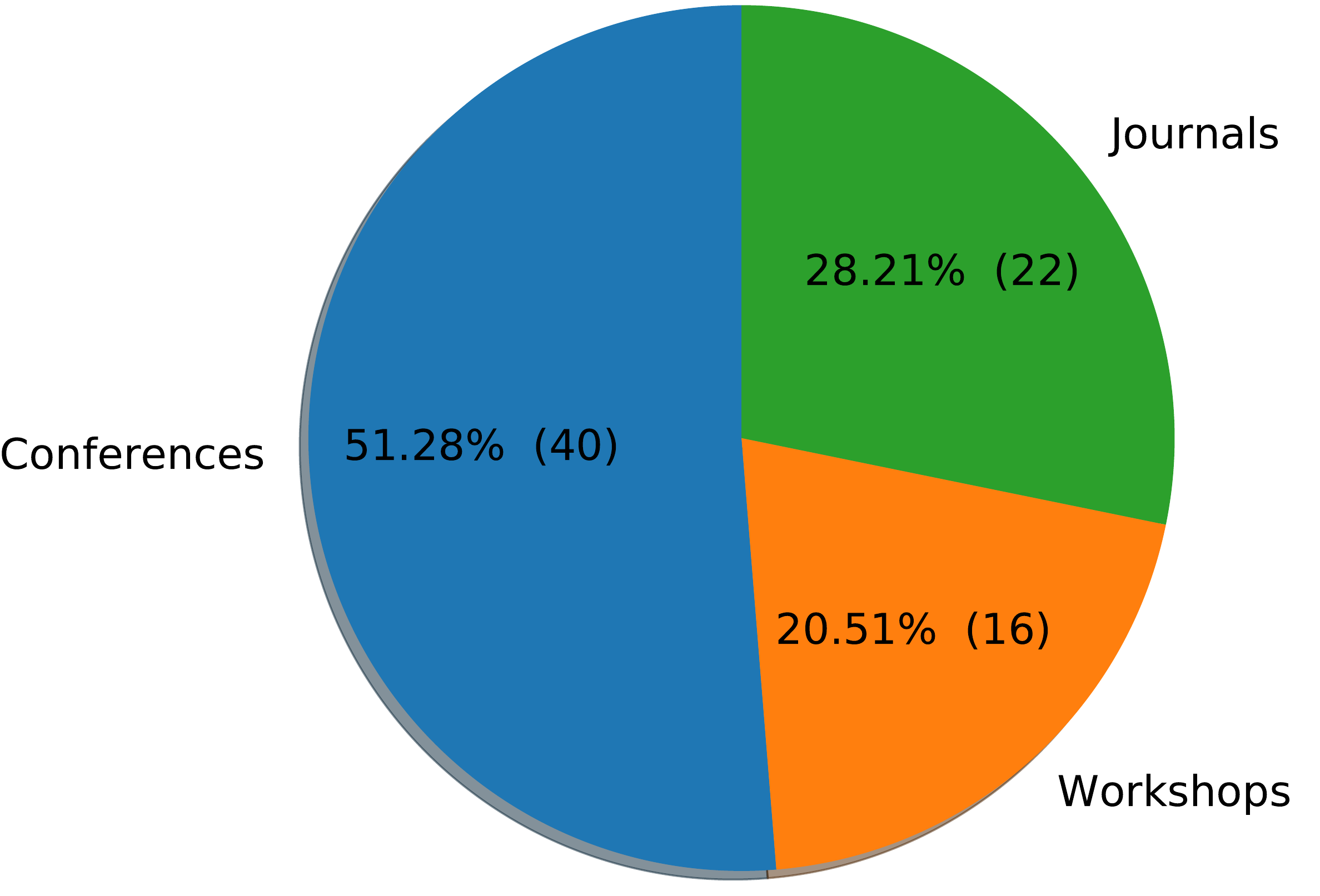}
\hfill
\includegraphics[width=.3\textwidth]{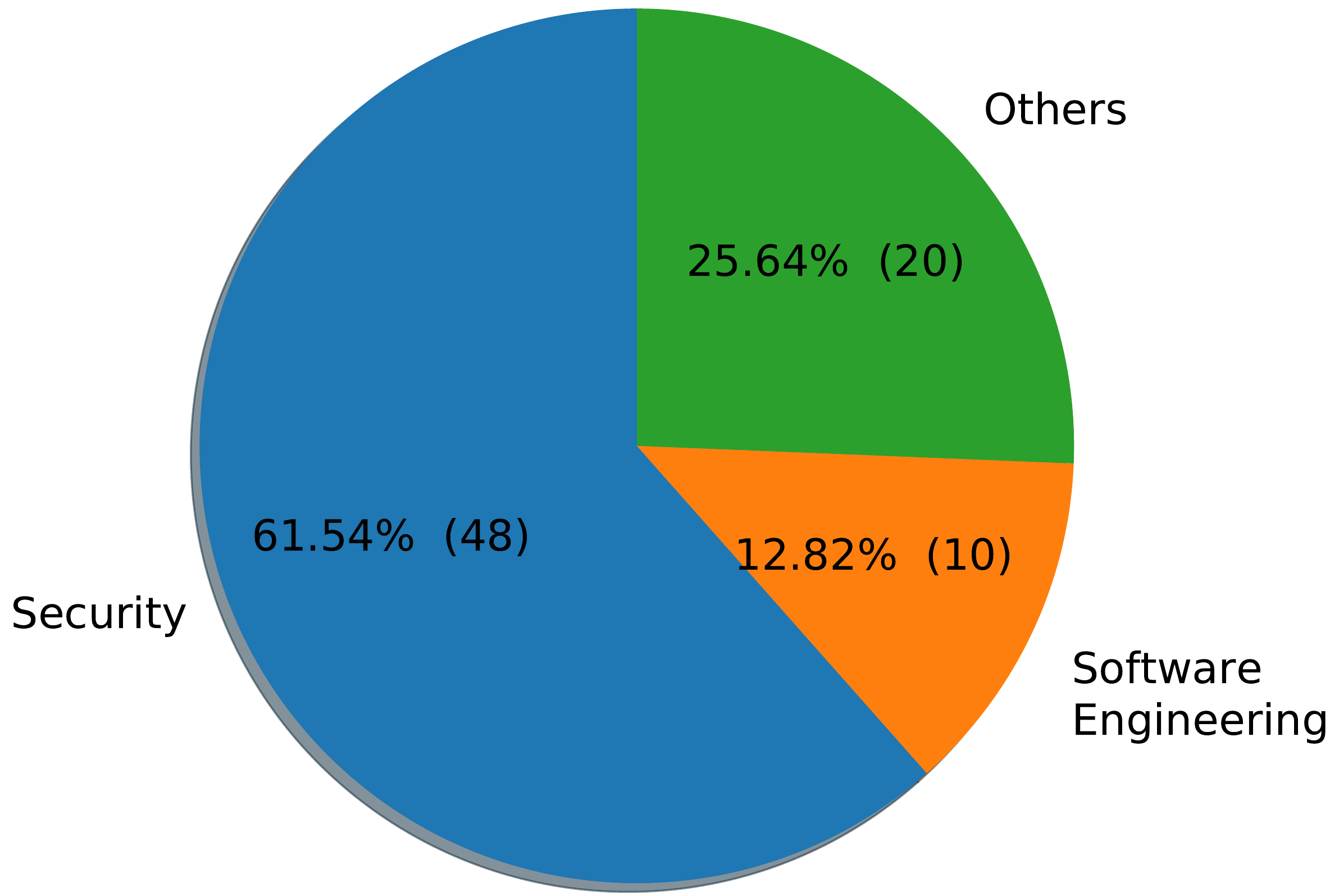}
\hfill
\includegraphics[width=.24\textwidth]{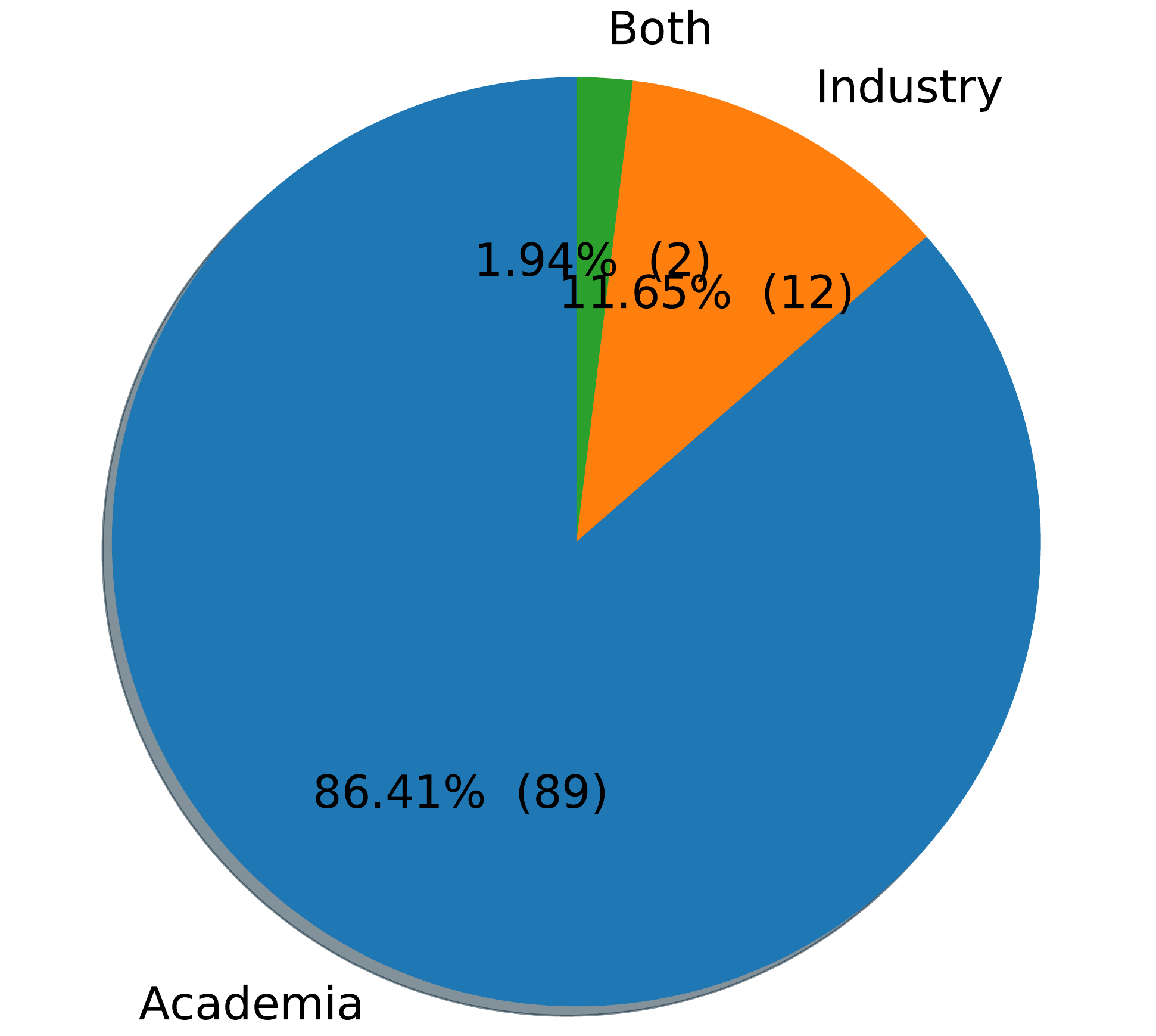}
\caption{\label{fig:pies} Distribution of papers by type of publication (left)
and field (middle). Distribution of authors by type of affiliation (right)}
\end{figure}

\begin{figure}
\centering
\includegraphics[width=\textwidth]{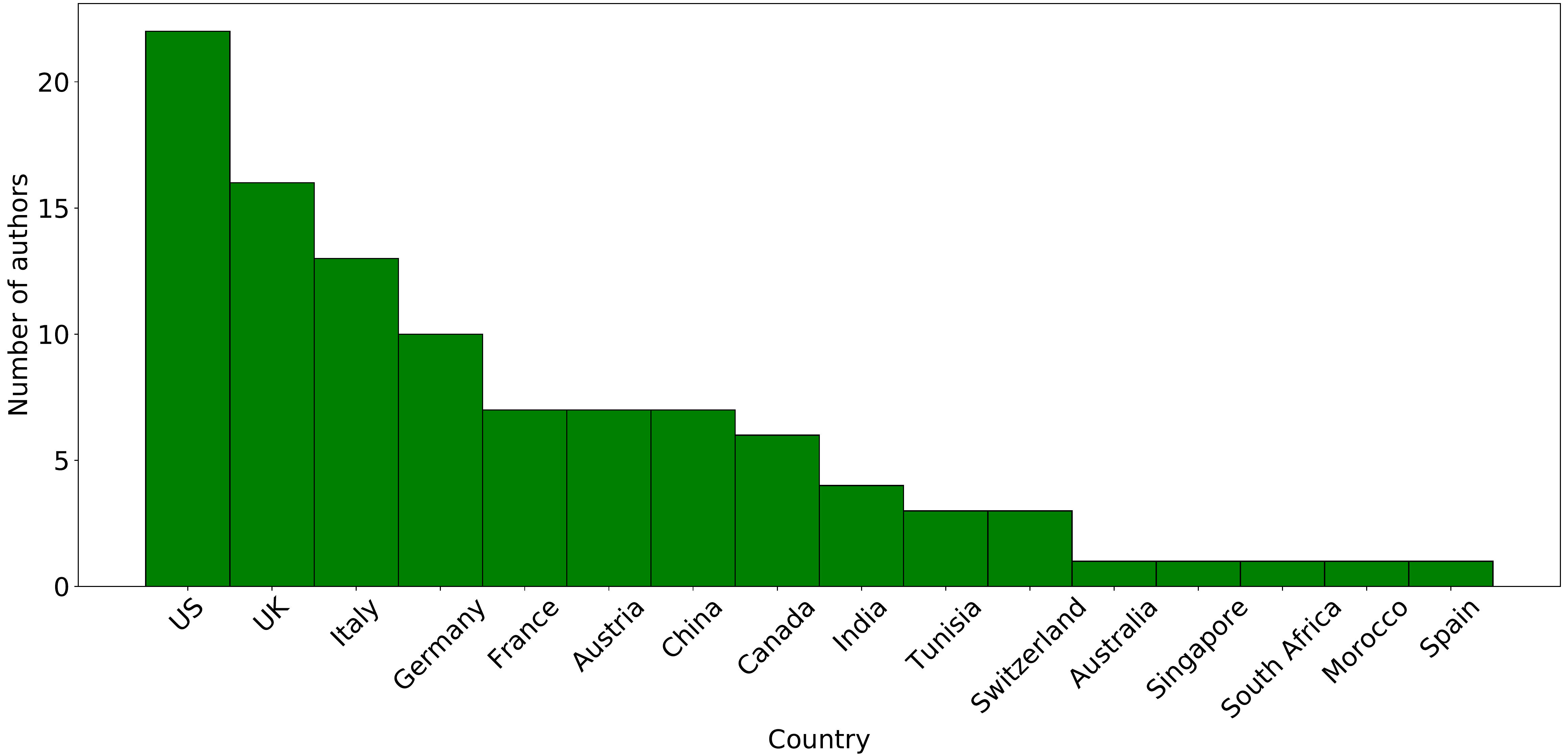}
\caption{\label{fig:countries} Distribution of the authors by country of
affiliation}
\end{figure}

Figure~\ref{fig:pies} shows, on the left, the distribution of papers by type of
publication (conference, workshop or journal); in the middle, the distribution
of papers by field of the venue where they were published (security, software
engineering, which includes BPM, and others, e.g., algorithms, theory, etc.);
and on the right, the distribution of authors by type of affiliation (academia,
industry or both). The pie charts show relative numbers and absolute numbers in
parentheses. As expected, most works are published in conferences (40 papers)
about security (48) by authors working in academia.
%%%%%%%%%%%%%%%%%%%%%%%%%%%%%%%%%%%%%%%%%%%%%%%%%%%%%%%%%%%%%%%%%%%%%%%%%%%%%%%%
%%%%%%%%%%%%%%%%%%%%%%%%%%%%%%%%%%%%%%%%%%%%%%%%%%%%%%%%%%%%%%%%%%%%%%%%%%%%%%%%
Figure~\ref{fig:countries} shows the distribution of authors by country of
affiliation. In Figures~\ref{fig:pies} and~\ref{fig:countries}, each author is
counted only once, even if he/she has published more than one paper. The total
number of authors is 103, which mostly work in academia (89), either in the
United States (22) or in Europe (56 authors from 6 countries).
%%%%%%%%%%%%%%%%%%%%%%%%%%%%%%%%%%%%%%%%%%%%%%%%%%%%%%%%%%%%%%%%%%%%%%%%%%%%%%%%
%%%%%%%%%%%%%%%%%%%%%%%%%%%%%%%%%%%%%%%%%%%%%%%%%%%%%%%%%%%%%%%%%%%%%%%%%%%%%%%%

\section{Workflows and authorization}
\label{sec:spec}

To understand why there are so many variants of the WSP and so many different
possible solutions, it is necessary to understand the relationship between
workflow models and authorization, as well as how security-sensitive workflows
are specified.

A workflow specification spans at least three perspectives: control-flow,
data-flow, and authorization (also called the resource
perspective)~\cite{aalst2003}. Control-flow constrains the execution order of
the tasks (e.g., sequential, parallel, or alternative execution); the
data-flow defines the various data objects consumed or produced by these tasks;
and the authorization specifies the organizational actors responsible for the
execution of the tasks in the form of authorization policies and constraints.
These three dimensions are interconnected, as each one of them influences the
others. The set of behaviors (i.e., possible executions of the workflow)
allowed by the control-flow is further constrained by conditions on the data,
as well as by user assignments and constraints in the authorization
perspective.

%%%%%%%%%%%%%%%%%%%%%%%%%%%%%%%%%%%%%%%%%%%%%%%%%%%%%%%%%%%%%%%%%%%%%%%%%%%%%%%%
\begin{figure}[t]
\centering
\includegraphics[width=\textwidth]{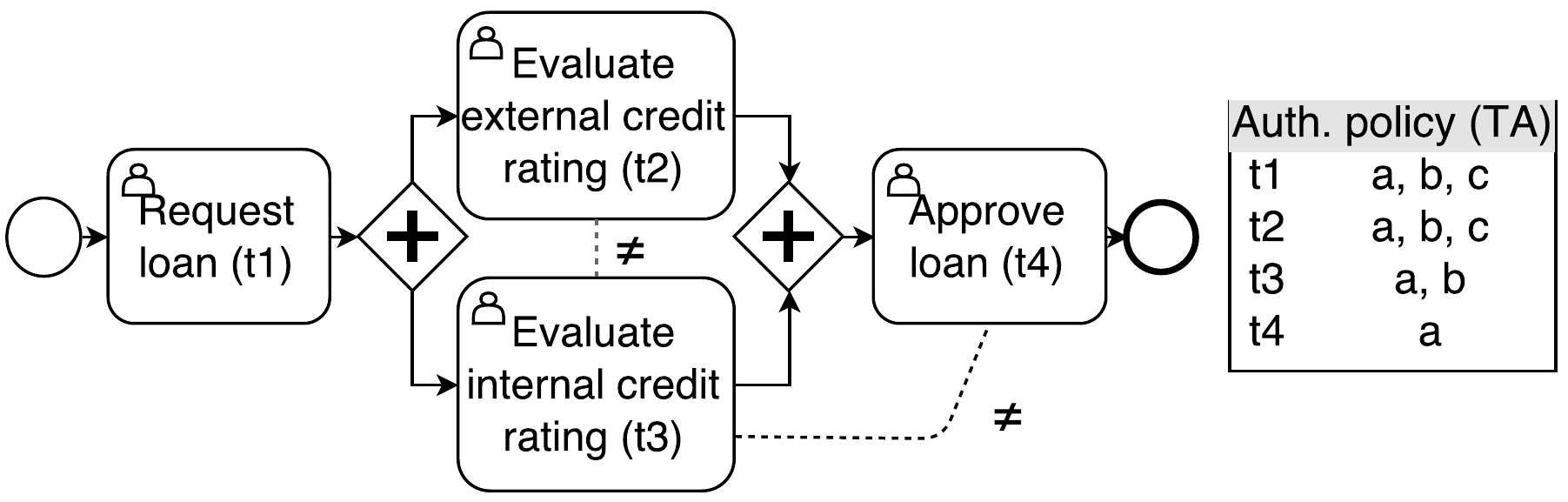}
\caption{\label{fig:example} Loan Origination Process in BPMN (left) and
authorization policy (right)}
\end{figure}
%%%%%%%%%%%%%%%%%%%%%%%%%%%%%%%%%%%%%%%%%%%%%%%%%%%%%%%%%%%%%%%%%%%%%%%%%%%%%%%%
Consider a simple Loan Origination Process with fours tasks (as shown in
Figure~\ref{fig:example} using the BPMN notation~\cite{weske2012}):
\emph{Request Loan} ($t1$), \emph{Evaluate External Credit Rating} ($t2$),
\emph{Evaluate Internal Credit Rating} ($t3$), and \emph{Approve Loan} ($t4$).
If task $t1$ has to be executed first, followed by $t2$ and $t3$ (in any
order), followed by $t4$, then the behaviors $t1,t2,t3,t4$ and $t1,t3,t2,t4$
are allowed, whereas, e.g., $t1,t4,t3,t2$ is not (where $t1,\dots,tn$
represents a sequence of $n$ tasks executed in order, i.e., $t_{i+1}$ is
executed after $t_i$). Now imagine that $t2$ is only executed for loans of more
than 10k Euro, then behavior $t1,t3,t4$ becomes allowed, but only for some
instances (those where the data object ``loan amount'' is less than 10k). If
the organization running this workflow adopts the policy shown on the fight of
Figure~\ref{fig:example} and the SoD constraints between $t2$ and $t3$ and
between $t3$ and $t4$ (shown as dashed lines labeled by $\neq$ in the Figure),
then any behavior containing, e.g., $t2(a)$ and $t3(a)$ is not allowed (where
$t(u)$ means that user $u$ executes task $t$).

Given the conflicting goals of business compliance and business continuity,
finding good (or even optimal) trade-offs has been a topic of research in the
Business Process Management (BPM) and security communities. The problems raised
by these opposing views are further complicated by the interplay between the
three perspectives (control-flow, data-flow, and authorization) introduced
above. Notice that a common practice in the analysis of workflow satisfiability
and resiliency is to abstract away from parts of a workflow specification. For
instance, few works take into account the data-flow (some completely disregard
it, e.g.~\cite{crampton2016}, and some model it with non-deterministic
decisions, e.g.,~\cite{basin2014}). It is also usual practice to limit the
allowed control-flow constructs and supported authorization constraints.
Different formulations of the WSP are concerned with at least three dimensions:
control-flow models, supported authorization constraints, and problem setting.

\subsection{Control-flow}
There are three basic categories of control-flow support: linear workflows,
which only admit a sequential execution of tasks
(e.g.,~\cite{bertino1999}); partial orders, which also allow parallel
executions (e.g.,~\cite{crampton2016}); and others (e.g.,
CSP~\cite{hoare1985} and Petri nets~\cite{murata1989}), which add support for
conditional branches and loops (e.g.,~\cite{basin2014}). It is known that a
family of partial orders is needed to characterize one Petri
net~\cite{katoen1995}, which means that modeling the control-flow with Petri
nets has the advantage of compactly representing a workflow that has to be
specified as potentially many partial orders. It is also always possible to
obtain a safe Petri net from a CSP process~\cite{winskel1987}. As described
in~\cite{wolter2008}, conditional execution can lead to execution paths of
different lengths, which means that WSP solutions that try to assign users to
every task in a workflow cannot be immediately applied.

\subsection{Authorization}
A plan $\pi : T \rightarrow U$, where $T$ is the finite set of tasks in the
workflow and $U$ is a finite set of users, is an assignment of tasks to users
representing a workflow execution where $(t,u) \in \pi$ means that user $u$
takes the responsibility of executing task $t$. An authorization constraint $c
\in C$ can be seen as a pair $(T', \Theta)$, where $T' \subseteq T$ is called
the scope of $c$ and $\Theta$ is a set of functions $\theta : T' \rightarrow
U$~\cite{crampton2016}. The functions in $\Theta$ specify the assignments of
tasks to users that satisfy the constraint. Instead of enumerating every
function $\theta \in \Theta$, it is common to define $\Theta$ implicitly by
using a specification device. Several classes of authorization constraints for
workflows have been identified in the literature. They can all be used, with
some ingenuity, to define the functions $\theta \in \Theta$, so they can be
recast in the form $(T',\Theta)$ shown above~\cite{cohen2014}.

\begin{description}
\item[Counting constraints] are of the form $(t_l,t_r,T')$, where $1 \leq t_l
\leq t_r \leq k$. A plan satisfies a counting constraint if a user performs
either no tasks in $T'$ or between $t_l$ and $t_r$ tasks. One example of
counting constraint is $(1,2,\{t1,t2,t3\})$, which is satisfied if a user
$u1$ executes $0$, $1$ or $2$ tasks among those in $\{t1,t2,t3\}$.

\item[Entailment constraints] are of the form $(T_1,T_2,\rho)$, where $T_1
\cup T_2 = T'$ and $\rho \subseteq U \times U$. A plan satisfies an
entailment constraint iff there exist $t_1 \in T_1$ and $t_2 \in T_2$ such
that $(\pi(t_1), \pi(t_2)) \in \rho$. In Type 1 constraints, both sets $T_1$
and $T_2$ are singletons. In Type 2 constraints, at least one of the sets
must be a singleton, whereas in Type 3 there are no restrictions on the
cardinality of sets. Examples of Type 1, 2 and 3 constraints are
$(\{t1\},\{t2\},\neq\})$, $(\{t1,t2\},\{t3\},\neq\})$, and
$(\{t1,t2\},\{t3,t4\},\neq\})$, respectively. The first constraint is
satisfied if a user $u1$ executes $t1$ and $u2$ executes $t2$ (because $u1
\neq u2$). The second and third constraints are satisfied if $u1$ executes
$t1$ and $u2$ executes $t3$. Those are examples of SoD constraints, BoD
constraints can be similarly defined by using $=$ instead of $\neq$. A
special class of Type 1 constraints are equivalence-based constraints, of the
form $(t_1,t_2,\sim)$, where $\sim$ is an equivalence relation on $U$. A plan
satisfies this kind of constraint if the user who executes $t_1$ and the user
who executes $t_2$ belong to the same equivalence class, e.g., same role (or
to different classes for $\not\sim$ constraints).

\item[User-independent constraints] $c$ are those where given a plan $\pi$
that satisfies $c$ and any permutation $\phi : U \rightarrow U$, the plan
$\pi' = \phi(\pi(s))$ also satisfies $c$~\cite{cohen2014}. I.e.,
user-independent constraints are those whose satisfaction does not depend on
the individual identities of users. The SoD constraints presented so far are
user-independent, whereas a constraint requiring a specific user to perform
at least one task in a set is not user-independent~\cite{cohen2016}.

\item[Class-independent constraints] are those whose satisfaction depends
only on the equivalence classes that users belong to~\cite{crampton2016}.
Formally, let $c$ be a constraint, $\sim$ be an equivalence relation on $U$,
$U^{\sim}$ be the set of equivalence classes induced by $\sim$, and $u^{\sim}
\in U^{\sim}$ be the equivalence class containing $u$. Then, for any plan
$\pi$, we can define a function $\pi^{\sim} : T \rightarrow U^{\sim}$ as
$\pi^{\sim}(t) = (\pi(t))^{\sim}$. Finally, $c$ is class-independent for
$\sim$ if for any function $\theta$, $\theta^{\sim} \in \Theta$ implies
$\theta \in \Theta$, and for any permutation $\phi: U^{\sim} \rightarrow
U^{\sim}$, $\theta^{\sim} \in \Theta^{\sim}$ implies $\phi \circ
\theta^{\sim} \in \Theta^{\sim}$~\cite{crampton2016}. One example of
class-independent constraint is $(\{t1\},\{t2\},\sim)$, where the classes
induced by $\sim$ corresponds to departments of a company. This constraint is
satisfied if $u(t1) \sim u(t2)$, i.e., the user executing $t1$ and the user
executing $t2$ are in the same department. Indeed, every equivalence
constraint $(t_1,t_2,\sim)$ (or $(t_1,t_2,\not\sim)$) is class-independent
and every user-independent constraint is class-independent with respect to
the identity relation~\cite{crampton2016}.
\end{description}

Other approaches to authorization constraint specification include Bertino et
al.'s constraint specification language~\cite{bertino1999} and Li and Wang's
Separation of Duties Algebra (SoDA)~\cite{li2008}. The first is based on rules
built on pre-defined logic predicates. The resulting set of rules, called
constraint base, is a stratified normal program. The second is an algebra for
high-level policies that allows to express and formalize policies based on
users' attributes and the number of users executing tasks. The policies are
enforced by low-level mechanisms such as static and dynamic separation of
duties in Role-Based Access Control (RBAC)~\cite{sandhu1996}.

Unlike for control-flow, it is not easy to classify authorization constraints
in terms of expressiveness, partly because there are many different frameworks
to express them. For instance, entailment constraints of Type 3 clearly include
those of Types 1 and 2, but counting constraints can also be used to express
some forms of SoD~\cite{wolter2008}, so entailment and counting constraints are
not disjoint (i.e., in some cases, it is possible to express the same set of
behaviors using a counting constraint or an entailment one). Also, clearly
user-independent and class-independent constraints subsume parts of the other
classes, but it is not clear which parts.

\begin{figure}[t]
\centering
\includegraphics{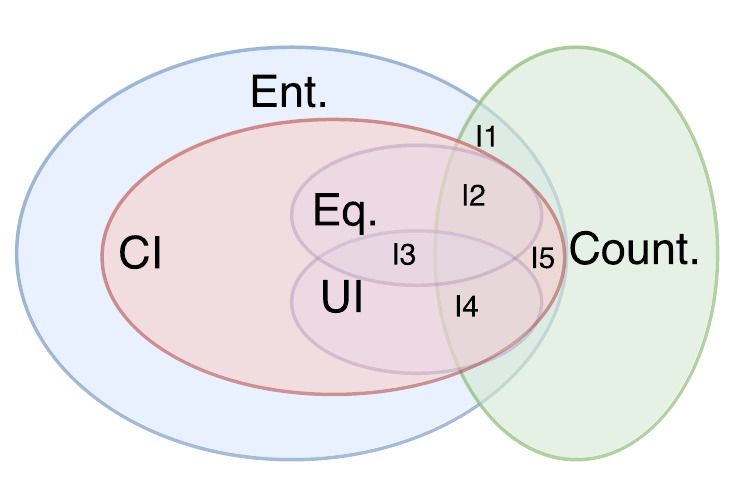}
\caption{\label{fig:cclasses} Constraint classes}
\end{figure}

Figure~\ref{fig:cclasses} shows an attempt to systematically classify some
classes of authorization constraints for workflow systems presented in the
literature.
%%%%%%%%%%%%%%%%%%%%%%%%%%%%%%%%%%%%%%%%%%%%%%%%%%%%%%%%%%%%%%%%%%%%%%%%%%%%%%%%
%%%%%%%%%%%%%%%%%%%%%%%%%%%%%%%%%%%%%%%%%%%%%%%%%%%%%%%%%%%%%%%%%%%%%%%%%%%%%%%%
The Figure shows the sets $\mathit{Ent.}$ of entailment constraints (the
subsets of constraints of Types 1, 2, and 3 are not shown to keep the figure
readable), $\mathit{Count.}$ of counting constraints, $\mathit{Eq.}$ of
equivalence constraints, $\mathit{CI}$ of class-independent constraints and
$\mathit{UI}$ of user-independent constraints. Naturally, $\mathit{Eq. \subset
Ent.}$ and $\mathit{CI. \subset Ent.}$, since an equivalence relation is an
instance of a binary relation. The facts $\mathit{UI} \subset \mathit{CI}$ and
$\mathit{Eq. \subset CI}$ were shown by Crampton et al.~\cite{crampton2016}.

The Figure also shows the following intersections:
$I1 = \mathit{Ent. \cap Count.}$,
$I2 = \mathit{Eq. \cap Count.}$,
$I3 = \mathit{Eq. \cap UI}$,
$I4 = \mathit{Count. \cap UI}$,
$I5 = \mathit{Count. \cap CI}$.
We can show that these intersections are non-empty by using SoD and BoD
constraints as examples. $I1$ and $I2$ are non-empty because SoD and BoD
can be specified using entailment: $(t1,t2,\neq)$ and $(t1,2,=)$, resp.;
counting: $(1,1,\{t1,t2\})$ and $(2,2,\{t1,t2\})$, resp.; or equivalence,
since $=$ is an equivalence relation. $I3$, $I4$, and $I5$ are non-empty
because both constraints are user-independent~\cite{cohen2016}, which also
makes them class-independent~\cite{crampton2016}.

To the best of our knowledge, there has never been a comparison between the
expressive power of other frameworks, e.g., SoDA and the constraint classes
defined by Crampton et al. In any case, the most widely adopted kinds of
constraints in practice (and in the examples of the works that we describe
below) are simple forms of SoD and BoD.

\subsection{Problem setting}
Different formulations of the WSP consider at least two distinguishing
characteristics: (i) is the order of the tasks considered? and (ii) is
satisfiability checked at design-time (before the execution of any instance of
the workflow) or at run-time (during execution)?

The separation between ordered and unordered WSP was presented
in~\cite{crampton2013}. The unordered WSP admits as solution a plan $\pi$
assigning users to tasks in such a way that all tasks have an assigned user and
all constraints are satisfied. The ordered version admits as solution a plan
$\pi$ with an execution schedule $\sigma$, which is a tuple $(t_1,\dots,t_k)$
such that $1 \leq i < j \leq k , t_i \neq t_j$, i.e., the assignment must
respect the ordering of tasks defined by the control-flow. The ordered and
unordered versions of the WSP are only equivalent for the class of well-formed
workflows~\cite{crampton2013}, i.e., workflows with the following property: for
all tasks $t_i, t_j$ that can be executed in any order, $(t_i,t_j,\rho) \in
C$ if and only if $(t_j,t_i,\tilde{\rho}) \in C$, where $\tilde{\rho}$ is
defined as $\{(u,u') \in U \times U : (u',u) \in \rho\}$ and $C$ is a set of
entailment constraints.

A classification of WSP approaches in the design-time/run-time dimension
was done in a recent survey~\cite{holderer2015}. Design-time techniques ensure
the existence of at least one satisfying assignment, whereas run-time
techniques enforce that a workflow instance follows a satisfying execution. As
shown in~\cite{dbsec2015}, it is possible to use, at run-time, an algorithm
that statically solves the WSP, but this is very inefficient, as it entails
solving a new instance of the problem for each user request.

\section{Workflow Satisfiability}
\label{sec:wsp}

Below, we describe the papers in the WSP category of Table~\ref{tab:papers}.
Each sub-category of Table~\ref{tab:papers} is reflected on a sub-section.
Table~\ref{tab:related} presents a comparison of relevant papers in each
category (column `Paper') in terms of control-flow models (column
`Control-flow'), authorization constraints (column `Constraints'), and problem
setting (ordered/unordered and execution time in columns `Ordered' and `Time',
respectively), as described in the previous Section.
%%%%%%%%%%%%%%%%%%%%%%%%%%%%%%%%%%%%%%%%%%%%%%%%%%%%%%%%%%%%%%%%%%%%%%%%%%%%%%%%
\begin{table}
\centering
\caption{\label{tab:related} Comparison of works in workflow satisfiability}
\begin{small}
\begin{tabular}{|c|c|c|c|c|}
\hline
Paper & Control-flow & Constraints & Ordered & Time \\\hline
\hline\multicolumn{5}{|c|}{Initial works}\\\hline\hline
\cite{bertino1999}
& Linear & Constraint Spec. Language  & Yes & Design-time \\\hline
\cite{crampton2005}
& P. order & Type 1 & Yes & Design-time \\\hline
\cite{wang2010}
& P. order & Type 2 & No & Design-time \\\hline
\hline\multicolumn{5}{|c|}{FPT algorithms}\\\hline\hline
\cite{crampton2013tissec}
& P. order & Type 3 + Count + Equiv. & No & Design-time \\\hline
\cite{crampton2013}
& P. order & Type 3 & No & Design-time \\\hline
\cite{cohen2014}
& P. order & User-independent + Equiv. & No & Design-time \\\hline
\cite{cohen2016}
& P. order & User-independent + Count & No & Design-time \\\hline
\cite{crampton2016}
& P. order & Class-independent & No & Design-time \\\hline
\cite{crampton2017}
& Conditional & User-independent & No & Design-time \\\hline
\hline\multicolumn{5}{|c|}{Model checking}\\\hline\hline
\cite{crampton2014}
& P. order & Type 1 & Yes & Design-time \\\hline
\cite{bertolissi2015}
& 1-safe PN & First-order logic & Yes & Run-time \\\hline
\hline\multicolumn{5}{|c|}{Others}\\\hline\hline
\cite{basin2014}
& CSP & SoD + BoD & Yes & Run-time \\\hline
\end{tabular}
\end{small}
\end{table}
%%%%%%%%%%%%%%%%%%%%%%%%%%%%%%%%%%%%%%%%%%%%%%%%%%%%%%%%%%%%%%%%%%%%%%%%%%%%%%%%
Notice that, for the sake of readability, only the most relevant papers
described in the next Sections are shown in the Table.

\subsection{Initial works}
The seminal works of Bertino et al.~\cite{bertino1997,bertino1999} described
the specification and enforcement of authorization constraints in workflow
management systems, presenting constraints as clauses in a logic program and an
exponential algorithm for assigning users and roles to tasks without violating
them, but considering only linear workflows. Tan et al.~\cite{tan2004} defined
a model for constrained workflow systems that includes constraints such as
cardinality, SoD and BoD. They defined the notion of a workflow specification
as a partial order on the set of tasks and of a constrained workflow
authorization schema, associating roles to tasks. Their main result is to find
conditions for the set of constraints that ensure that for any user authorized
to a task, there is at least one complete workflow instance when this user
executes this task. Crampton~\cite{crampton2005} extended these ideas by
defining Type 1 constraints, and developing an algorithm to determine whether
there exists an assignment of users to tasks that satisfies the constraints.
Solworth~\cite{solworth2006} defined an approvability graph to describe
sequences of actions defining the termination of workflows with an RBAC policy,
linear or conditional executions and the possibility of loops. In the same
work, the author shows a polynomial algorithm to determine the minimum number
of users per role to ensure that a workflow can terminate.

Wang and Li~\cite{wang2007,wang2010} introduced the unordered version of the
WSP, showed that it is NP-complete and that this intractability is inherent in
authorization systems supporting simple constraints. They reduced the problem
to SAT, which allows the use of off-the-shelf solvers, and showed that, with
only equality and inequality relations (BoD/SoD), the WSP is Fixed-Parameter
Tractable (FPT) in the number of tasks (since the number of tasks is typically
smaller than the number of users)\footnote{FPT is a parameterized complexity
  class which contains the problems that can be solved in time $f(k)\cdot n^a$
  for some computable function $f$, parameter $k$, and constant
  $a$~\cite{downey2013}.
  Many hard problems become less complex if some natural parameter of the
  instance is bounded. An example is the satisfiability problem parameterized
  by the number of variables: a given formula of size $n$ with $k$ variables
  can be checked by brute force in time $O(2^kn)$. The WSP is FPT
  when parameterized by the number of tasks (i.e., $k = |T|$).}.
Wang and Li's FPT proof motivated many later works by Crampton et al., mostly
considering the unordered version of the WSP for workflows specified as
partial orders.

\subsection{FPT algorithms}
Crampton et al.~\cite{crampton2012,crampton2013tissec} improved the complexity
bounds for the WSP and showed that it remains FPT with counting and equivalence
constraints. Later~\cite{crampton2013}, they used the notion of constraint
expressions (logical combinations of constraints) to support conditional
workflows and Type 3 constraints by essentially splitting one instance of the
problem into many instances, e.g., an instance of WSP for SoD/BoD constraints
of Type 3 can be transformed into multiple instances of the WSP with SoD/BoD
constraints of Type 1, and an instance of the WSP for a conditional workflow
can be solved as many instances for parallel workflows. They also showed that
the ordered version of the WSP is FPT for constraints of Type 1.
In~\cite{crampton2013faw}, they showed that the WSP remains FPT with seniority
constraints.

Crampton et al. first presented FPT algorithms for the WSP with
user-independent constraints in~\cite{cohen2014faw} and improved them
in~\cite{karapetyan2015} with pattern backtracking. Cohen et
al.~\cite{cohen2014} solved the WSP using techniques for the Constraint
Satisfaction Problem, which allowed the authors to devise a general algorithm
that works for several families of constraints. Their solution builds
executions incrementally, discarding partial executions that can never satisfy
the constraints. The authors showed that their algorithm is optimal for
user-independent constraints. Cohen et al.~\cite{cohen2016} demonstrated the
practicality of the previously designed algorithm by adapting it to the class
of user-independent counting constraints and showing its superiority when
compared with the classical SAT reduction of the problem. Gutin et
al.~\cite{gutin2016,gutin2016ipl} studied the kernelization\footnote{A
  kernelization of a parameterized problem $\mathcal{Q}$ is a polynomial-time
  computable function $K: (x,k) \mapsto (x',k')$ such that $(x, k) \in
  \mathcal{Q}$ iff $(x', k') \in \mathcal{Q}$, and such that $|x'|, k' \leq
  h(k)$ for some $h(k)$. Here, $(x, k)$ is an instance of $\mathcal{Q}$, and
  $h(k)$ is the size of the kernel.}
of the WSP and demonstrated that $O^*(2^{k \log_2 k})$ is a tight lower bound
for the WSP with user-independent constraints (assuming the Strong Exponential
Time Hypothesis).

Crampton et al.~\cite{crampton2015ipec,crampton2016} extended the notion of
user-independent constraints to that of class-independent constraints, showed
that the WSP remains FPT in this case and provided an algorithm to solve it.
Crampton et al.~\cite{crampton2016} and Cohen et al.~\cite{cohen2016}
experimentally compared the results of FPT algorithms against those of a SAT
solver on workflows of up to $30$ tasks and concluded that FPT algorithms are
better because those based on the SAT solver run out of memory. Finally,
Crampton et al.~\cite{crampton2017} extended the applicability of their FPT
solution to support conditional workflows with release points, which specify
that a constraint may be active only for some execution branches; as
in~\cite{crampton2013}, the solution is to split a workflow instance into
many ones and solve multiple instances of the WSP.

\subsection{Model checking}
Xiangpeng et al.~\cite{xiangpeng2006} presented a framework to integrate RBAC
into the Business Process Execution Language (BPEL)~\cite{weske2012}, with
authorization constraints expressed in temporal logic. They used model checking
to verify that a given BPEL process satisfies its security constraints. Wolter
et al.~\cite{wolter2009} presented an approach to verify security properties of
an annotated business process model by automatically translating it into a
process meta language and using SPIN\footnote{\url{http://spinroot.com/}} for
verification. Their implementation was integrated as a plug-in for the modeling
tool Oryx\footnote{\url{http://bpt.hpi.uni-potsdam.de/Oryx}}.

Crampton and Huth~\cite{huth2011,crampton2012bpmw} showed that model checking
on an NP-complete fragment of Linear Temporal Logic (LTL), called LTL(F), can
be used to create and validate plans for security-sensitive workflows and
argued that this approach is more robust, uniform, and expressive than previous
formalizations. Later~\cite{crampton2012paar}, they investigated a
propositional encoding with Binary Decision Diagrams (BDDs) and compared it
with the model checking approach to guarantee the satisfiability of workflow
instances statically or dynamically. They showed that the propositional
encoding is too costly because it has to be called for each access request at
run-time, whereas the model checking approach can pre-compute a set of
solutions for a workflow instance. In yet another work, Crampton et
al.~\cite{crampton2014} presented three encodings in LTL(F) that can compute a
set of solutions to the WSP. The slowest encoding considers the ordered WSP,
while the other two consider unordered versions. They experimented with
workflows of up to $220$ tasks. The synthesis of monitors was left as future
work.

Bertolissi and Ranise~\cite{bertolissi2013} introduced a class of symbolic
transition systems, called composed array-based systems, capable of
representing collections of security-sensitive workflows. They studied the
verification of reachability properties of such systems and found sufficient
conditions for the termination of a reachability analysis procedure.
Later~\cite{bertolissi2013icitst,bertolissi2014}, the same authors used this
class of systems and proposed a methodology based on Satisfiability Modulo
Theories (SMT) solving~\cite{barrett2009} to build run-time monitors capable of
ensuring the successful termination of workflows subject to authorization
constraints. This methodology was extended in~\cite{bertolissi2015} with a
fully automated technique, an implementation, and an experimental evaluation.
The technique works by synthesizing run-time monitors capable of ensuring that
all executions terminate and authorization constraints in a workflow are
satisfied. In~\cite{sacmat2016} an extension of this technique was described.
It is based on a refinement of the transition systems used to specify
security-sensitive workflows. The refined transition systems are associated to
a suitable notion of interface, forming a so-called security-sensitive
component. The authors then show how to synthesize monitors for components and
how to combine these monitors in a principled way.

%%%%%%%%%%%%%%%%%%%%%%%%%%%%%%%%%%%%%%%%%%%%%%%%%%%%%%%%%%%%%%%%%%%%%%%%%%%%%%%%
%In~\cite{sefm2017}, the authors showed how to encode more complex
%authorization constraints in the declarative input language of the SMT-based
%model checker, therefore expanding the scope of applicability of their
%techniques.
%%%%%%%%%%%%%%%%%%%%%%%%%%%%%%%%%%%%%%%%%%%%%%%%%%%%%%%%%%%%%%%%%%%%%%%%%%%%%%%%

This approach was implemented in two tools. The first, called
\textsc{Cerberus}~\cite{tacas2016}, integrates constraint specification,
monitor synthesis, and run-time enforcement in workflow management system. The
second, called \textsc{Aegis}~\cite{codaspy2017} can synthesize run-time
monitors for workflow-driven web applications. The monitors are capable of
enforcing control-flow and data-flow integrity, authorization policies and
constraints, as well as ensuring the termination of workflows.

\subsection{Others}

Basin et al.~\cite{basin2012-sacmat} considered the problem of choosing
authorization policies that allow a successful workflow execution and an
optimal balance between system protection and user empowerment. They treated
the problem as an optimization problem (finding the cost-minimizing
authorization policy that allows a successful workflow execution) and showed
that, in the role-based case, it is NP-complete. They generalized the decision
problem of whether a given authorization policy allows a successful workflow
execution to the notion of an optimal authorization policy that satisfies this
property.

Burri and Karjoth~\cite{burri2012} studied the flexible scoping of
authorization constraints in workflows containing loops. They introduced the
notion of release points, which remove associations between users and their
previously executed tasks, and extended the Business Process Model Notation
(BPMN)~\cite{weske2012} to support it.

In a following work~\cite{basin2012}, the authors used SoDA to enforce
SoD constraints in a dynamic, service-oriented enterprise environment. They
generalized SoDA's semantics to workflow traces and refined it for control-flow
and role-based authorizations. Their formalization, based on CSP, is the base
for provisioning SoD as a Service, with an implementation using a workflow
engine and a SoD enforcement monitor. Finally, in~\cite{basin2011,basin2014},
they used CSP to model workflows in two levels: control-flow and task
execution, allowing them to synthesize monitors that enforce at run-time
obstruction-free, or satisfying, workflow executions. However, the monitor
in~\cite{basin2012} only verifies if a trace of a workflow satisfies a SoDA
term with respect to the past, being incapable of checking whether there is a
future trace that can be concatenated in order to satisfy the workflow. On the
other hand, the monitor in~\cite{basin2014} enforces obstruction-freedom (which
is equivalent to solving the WSP) but in an approximated way and may be too
restrictive. The authors call their monitors Enforcement Processes and the
problem of deciding the existence of such a monitor for a constrained workflow
is called Enforcement Process Existence (EPE). Their naive solution to the EPE
is double exponential in the number of users and constraints because it depends
on checking failure-equivalence in CSP~\cite{roscoe1994}. They present two
approximate solution, one is exponential and one is polynomial. The
approximations are based on solutions to the graph coloring
problem~\cite{chartrand2008} and are overly restrictive because they may return
`No' even if an enforcement process does exist for the constrained workflow
taken as input (although they make no approximations in the other direction,
i.e., if there does not exist an enforcement process, the procedures always
return `No'). They also implemented tool support for the specification and
enforcement of constraints, but the evaluation was limited to a few workflows
used as examples.

Schefer et al.~\cite{schefer2012} focused on BoD constraints in workflows with
RBAC. They categorized these constraints into subject-binding and role-binding
and provided algorithms to check their satisfiability. Yang et
al.~\cite{yang2013} defined several formulations of the WSP, considering
different control-flow patterns, studied their complexity, and showed that, in
general, the problem is intractable. Bo et al.~\cite{bo2016} proposed a method
to solve the WSP without exploring the space of all possible user-task
assignments. Their method is based on an Improved Separation of Duties Algebra
(ISoDA) to describe a WSP instance, which is reduced to multi-mutual-exclusion
expressions, whose satisfiability is determined by a bespoke algorithm.

Hummer et al.~\cite{hummer2013} studied the specification and enforcement of
entailment constraints in distributed business processes, and how to
detect---but not avoid---deadlocks. Ayed et al.~\cite{ayed2008,ayed2009}
considered the WSP and its complexity in distributed workflows, which can be
intra- or inter-organizational. They used Petri nets to model the control-flow
and Organization-Based Access Control (OrBAC)~\cite{elkalam2003} for
authorizations. Their approach starts from a global policy and derives local
policies that can be enforced by components running parts of a workflow.

Jemel et al.~\cite{jemel2015} presented how the ECA (Event-Condition-Action)
paradigm and agent technology can be exploited to steer authorization plans in
order to satisfy dynamic constraints. Especially, they studied inter-instance
constraints, i.e., constraints across instances of execution of the same
workflow model. Inter-instance authorization constraints and their
satisfiability in workflow systems have also been studied in~\cite{warner2006},
where the authors define a constraint specification language and propose
methodologies to identify cases in which SoD constraints result in an anomaly
(unsatisfiability is one of the identifiable anomalies, besides inconsistency
and overlaps). These anomalies can be detected at design-time and prevented at
run-time.

Holderer et al.~\cite{holderer2016} presented a hybrid solution to complete
workflow instances whose execution was stopped because of the constraints. If a
log is provided, they partition its traces into ``successful'' and
``obstructed'' by analyzing the given workflow and its authorizations. An
obstruction can then be solved by finding its nearest match in the successful
traces. If no log is provided, they flatten the workflow and its authorizations
into a Petri net and encode the obstruction with a corresponding ``obstruction
marking''. The structural theory of Petri nets is used to provide a vector that
may violate some firing rules, but reaches a final marking that completes the
workflow.

Combi et al.~\cite{combi2016} studied workflows with Temporal RBAC and the
satisfaction of temporal constraints for these workflows. They comment that
there are currently no approaches to workflow satisfiability and resiliency
that take into account temporal constraints and leave this as future work.

\section{Workflow Resiliency}
\label{sec:res}

%%%%%%%%%%%%%%%%%%%%%%%%%%%%%%%%%%%%%%%%%%%%%%%%%%%%%%%%%%%%%%%%%%%%%%%%%%%%%%%%
%%Below, we describe the papers in the resiliency category of
%%Table~\ref{tab:papers}. Each sub-category of Table~\ref{tab:papers} is
%%reflected on a paragraph. Table~\ref{tab:related-res} presents a summary of
%%the most relevant works solving the workflow resiliency problem, classified
%%as the works on the WSP by control-flow, supported constraints, and problem
%%setting.
%%%%%%%%%%%%%%%%%%%%%%%%%%%%%%%%%%%%%%%%%%%%%%%%%%%%%%%%%%%%%%%%%%%%%%%%%%%%%%%%
%%\begin{table}
%%\centering
%%\caption{\label{tab:related-res} Comparison of works in workflow resiliency}
%%\begin{tabular}{|c|c|c|c|c|}
%%\hline
%%Paper & Control-flow & Constraints & Ordered & Resiliency \\\hline\hline
%%\cite{li2009} & P. order & Type 2 & No &
%%\begin{tabular}{@{}c@{}} static \\ decremental \\ dynamic\end{tabular}
%%\\\hline
%%\cite{jcs2017} & 1-safe PN & First-order logic & Yes & static \\\hline
%%\cite{mace2014} & P. order & SoD + BoD & No & quantitative \\\hline
%%\cite{mace2015} & Task structure & SoD + BoD & No & quantitative \\\hline
%%\cite{crampton2017jcs} & P. order & User-independent & No &
%%quantitative \\\hline
%%\end{tabular}
%%\end{table}
%%%%%%%%%%%%%%%%%%%%%%%%%%%%%%%%%%%%%%%%%%%%%%%%%%%%%%%%%%%%%%%%%%%%%%%%%%%%%%%%

\paragraph{Static, dynamic}
Li et al.~\cite{li2009} introduced the notion of resiliency policies for access
control systems, i.e., policies that require the system to be resilient to the
absence of users. They defined the Resiliency Checking Problem (RCP), which
amounts to checking if an access control state satisfies a given resiliency
policy. Wang and Li~\cite{wang2010} then studied resiliency in workflow systems
and its relation to the WSP, defining three levels of resiliency based on when
the users are allowed to be absent and whether they are allowed to return. In
\emph{static} resiliency, a number of users may be absent before a workflow
instance execution; in \emph{decremental} resiliency, users may be absent
before or during a workflow instance execution, but absent users do not become
available again; and in \emph{dynamic} resiliency, users may become absent and
available again. They showed that checking static workflow resiliency is in NP,
while checking decremental and dynamic resiliency is in PSPACE. The authors
observed that there are other possible formulations of resiliency that can be
of interest.

The solution to workflow resiliency described in~\cite{jcs2017}, relies on
pre-computing reachability graphs~\cite{bertolissi2015} with a model checker
and refining them with a given authorization policy. The refinement is
performed by a depth-first search of the graph to prune those executions that
do not satisfy the authorization policy used in the deployment context under
consideration. This is combined with a (heuristic) method to generate subsets
of users not containing $k$ users (by adapting the pruning strategy
from~\cite{li2009}) in order to find scenarios guaranteeing the termination of
a workflow despite the absence of $k$ users. The authors only consider static
workflow resiliency, but they note that the solution could be adapted for
quantitative resiliency by assigning weights representing availability to the
edges in the reachability graph, as hinted at in~\cite{crampton2016sacmat}.
This work is an extension of~\cite{dbsec2015}, where the authors define a class
of Scenario Finding Problems, i.e. finding WSP solutions that also satisfy
other properties defined by the user (e.g., a particular user executing a task
or a minimal number of users). Each of these problems is solved by a different
refinement of a reachability graph.

%%%%%%%%%%%%%%%%%%%%%%%%%%%%%%%%%%%%%%%%%%%%%%%%%%%%%%%%%%%%%%%%%%%%%%%%%%%%%%%%
%%Crampton et al.~\cite{2017ciac} introduced a framework in parameterized
%%algorithms to solve resiliency versions of decision problems (including
%%workflow resiliency). They defined resiliency for Integer Linear Programs
%%(ILP) and proved that ILP Resiliency is FPT under a certain parameterization.
%%%%%%%%%%%%%%%%%%%%%%%%%%%%%%%%%%%%%%%%%%%%%%%%%%%%%%%%%%%%%%%%%%%%%%%%%%%%%%%%

\paragraph{Quantitative}
Mace et al.~\cite{mace2014} defined \emph{quantitative} workflow resiliency, in
which a user wants to know how likely a workflow instance is to terminate given
a user availability model. The authors solve the problem by finding optimal
plans for Markov Decision Processes (MDP). The same authors~\cite{mace2015}
showed that alternative executions may lead to different resiliency values for
each path, and defined resiliency variance as a metric to indicate volatility,
claiming that a higher variance increases the likelihood of workflow failure.
User availability models were discussed in more details
in~\cite{mace2015modelling}, categorized into non-deterministic, probabilistic,
and bounded, with several encodings for the PRISM probabilistic model
checker\footnote{\url{http://www.prismmodelchecker.org/}}. The same group
studied the impact of policy design (adding or removing authorization
constraints) on workflow resiliency computation time~\cite{mace2015impact}.
They were able to compute sets of security constraints that can be added to a
policy in order to reduce computation time while maintaining resiliency. The
authors then developed WRAD~\cite{mace2016}, a tool for workflow resiliency
analysis and design, which automatically encodes workflows into PRISM,
evaluates their resiliency and computes optimal changes for security
constraints to ensure a resiliency threshold.

Crampton et al.~\cite{crampton2017jcs} studied the Bi-Objective
WSP, which is the problem of minimizing two weight functions associated to
a valid plan, one representing the violation of constraints and one
representing the violation of the authorization policy. The authors related
this problem to workflow resiliency, claiming that Mace et al.'s translation to
MDP is not necessary since the same metrics can be computed by constructing a
graph where the nodes are partial valid plans, and the edges, connecting
successive plans, are labeled with the probability of a user being available to
execute the next task (checking every possible partial plan has
exponential-time complexity). The Bi-Objective WSP is a generalization of the
Valued WSP~\cite{crampton2015}, which has as single objective minimizing the
sum of both weights. Crampton et al.~\cite{crampton2016sacmat} reduced the RCP
to the WSP and showed how to solve it using an FPT algorithm for the WSP. The
RCP differs from workflow resiliency by considering three parameters: $s$
users, forming $d$ teams of size $t$, such that all teams are authorized to
access the resources in a policy $P$. In contrast, $k$-resiliency for workflows
just considers $k$ absent users and whether the remaining users can execute all
tasks. However, the RCP is always static. The basic solution in the original
paper about RCP~\cite{li2009} is to enumerate all subsets of $s$ users and
check for satisfiability (using a procedure for $s=0$ as a black-box), but
there is a pruning strategy based on the redundancy of some subsets, to have a
more efficient solution. Crampton et al.~\cite{crampton2016sacmat} solved the
same RCP problem, using the same pruning strategy and their FPT algorithm as
the black-box to decide satisfiability (by translating the resources in $P$ to
a workflow). The authors mention that this basic reduction cannot be applied
directly to decremental or dynamic workflow resiliency, but they point to their
work on Valued WSP~\cite{crampton2015} as a possibility to do it, by using
weights to represent the availability of users.

\paragraph{Others}
Paci et al.~\cite{paci2008} investigated resiliency in business processes
specified in RBAC-WS-BPEL (an extension of BPEL that supports the specification
of authorization policies and constraints). They extended RBAC-WS-BPEL with
resiliency constraints and the notion of user failure resiliency, then proposed
an algorithm to determine if a WS-BPEL process is user failure resilient.
Lowalekar et al.~\cite{lowalekar2009} proposed a quadratic programming
algorithm to generate user-task assignments that respect a security policy and
are statically resilient. Lu et al.~\cite{lu2014} studied dynamic workflow
adjustment, i.e., how to minimally adjust existing user-task assignments, when
a sudden change occurs, e.g., absence of users, so that the workflow can still
be satisfied.

\section{Related problems}
\label{sec:other}

Besides satisfiability and resiliency, there are other problems of interest to
users designing an authorization policy or deploying a workflow in their
organization. In this Section, we explore these related problems that have been
identified in the literature.

\paragraph{Workflow feasibility}
Khan and Fong~\cite{khan2012} defined the problem of workflow feasibility, when
there are rules to update the authorization relation (e.g., Administrative
RBAC~\cite{sandhu1996}). A workflow is feasible if there is at least one
reachable access control configuration where the workflow is satisfiable.
Later, Mehregan and Fong~\cite{mehregan2016} adapted the notions of workflow
satisfiability, resiliency, and feasibility to define a protocol for policy
negotiation on Relationship-Based Access Control (ReBAC)~\cite{fong2011} with
multiple ownership.

\paragraph{WSP with delegation}
One way of avoiding the problems caused by absent users is to use delegation, a
mechanism by which a user (delegator) can share or transfer a subset of his/her
permissions with another user (delegatee), so that, e.g., absent users can
transfer their permissions to available users in order not to disrupt the
execution of workflows. There are several models for delegation in access
control and Crampton and Khambhammettu~\cite{crampton2008,crampton2008sac}
discuss some models of delegation on workflow systems, depending on: execution
model, task type, and delegation type.
Crampton and Khambhammettu~\cite{crampton2008} described the relation between
authorization delegation and workflow satisfiability, proposing algorithms for
evaluating delegation requests in different workflow execution models, such
that the requests are granted if the workflows remain satisfiable after the
delegation. Crampton and Morisset~\cite{crampton2010} discussed an
auto-delegation mechanism to automatically respond to the absence of authorized
users, which they claim is useful in systems where the set of authorized users
changes unpredictably over time. El Bakkali~\cite{el2012,el2013} presented a
solution to bypass WSP deadlocks at run-time and enhance the resiliency of
workflow systems by using delegation and priority concepts.

\paragraph{Cardinality-Constrained Minimum User Problem}
The Cardinality-Constrained Minimum User Problem (CMUP)~\cite{roy2015} consists
in finding the minimum number of users required for a satisfiable workflow
instance. The original solutions to the CMUP apply either an integer
programming solver or an algorithm based on a generalization of a greedy
heuristic for coloring hypergraphs. Crampton et al.~\cite{crampton2017jcs}
showed that CMUP can be reduced to multiple instances of the WSP or a single
instance of the BO-WSP. Kohler and Schaad~\cite{kohler2008} proposed a
graph-based technique to compute minimal user bases to help policy designers
avoid such situations. Their work is limited to RBAC policies. dos Santos et
al.~\cite{dbsec2015,jcs2017} can compute minimal user bases by refining a
reachability graph (as done for the resiliency problem) and Crampton et
al.~\cite{crampton2014} can do the same with one of their encodings for LTL.

\paragraph{Purpose}
Jafari et al.~\cite{jafari2014} developed a framework for formalizing and
enforcing purpose-based privacy policies. They propose a language for
expressing constraints about purposes of actions, whose semantics are defined
over an abstract model of activities directly derivable from business
processes. They show how to tie purpose-based constraints and authorization
policies and present a model checking algorithm for verifying whether a state
of the system complies with a set of policies. This algorithm is then used in a
purpose reference monitor. The authors then comment that is possible to define
a workflow satisfiability problem which takes into account authorization and
purpose constraints, but this is left as future work. De Masellis et
al.~\cite{demasellis2015} proposed a declarative framework based on a
first-order temporal logic that allowed them to give a precise semantics to
purpose-aware policies and to reuse algorithms for the design of a run-time
monitor enforcing purpose-aware policies. They also observe that handling
purposes in the presence of authorization constraints requires to solve, at
run-time, the WSP.

\paragraph{Access control policy properties}
Sun et al.~\cite{sun2011} analyzed the complexity of authorization in RBAC with
security constraints by studying fundamental problems related to access control
constraints and user-role assignment. They developed algorithms and complexity
results for a series of problems and related them to the WSP. Intuitively, one
can map each role in their setting to one step in a workflow, and the problem
of assigning users to roles becomes the same as assigning users to steps.
Ranise and Traverso defined an Action Language for Policy Specification
(ALPS)~\cite{ranise2014} and used it to model access control for workflows.
They showed that instances of the WSP fall in the larger category of
reachability problems supported by their language. Garrison et
al.~\cite{garrison2014} formalized the access control suitability analysis
problem, which seeks to evaluate the degree to which a set of candidate access
control schemes can meet the needs of an application-specific workload. Part of
the solution to their problem is done via simulation, and during a simulation,
they have to solve instances of the WSP. Calzavara et al.~\cite{calzavara2016}
studied the problem of detecting collusion attacks on workflows with
administrative RBAC. Part of their static analysis technique amounts to solving
instances of the WSP. Berge et al.~\cite{berge2016} defined the class of
Authorization Policy Existence Problems (APEP), where a positive answer means
that an organization's objectives can be realized. The WSP and the Bi-Objective
WSP are both instances of an APEP. They also analyzed the complexity of these
problems and, for particular sub-classes of constraints, developed FPT
algorithms to solve them.

\section{Conclusion}
\label{sec:conclusion}

This survey investigated problems related to the satisfiability of
authorization policies in workflow systems. We did not include other very broad
and related areas, such as compliance, in order to limit the scope of the
literature review. We examined 78 publications between the years of 1997 and
2017 and categorized them based on problems and approaches to solve them.

We found out that the area is more active than ever, that there are many
variations on the basic WSP (e.g., ordered/unordered, design-time/run-time,
valued), that the study of this problem has led to other interesting questions
(e.g., resiliency and feasibility), and that some of these questions are even
applicable outside the context of workflow management systems (e.g., web
applications~\cite{codaspy2017}). We can answer the three research questions
posed at the beginning as follows.
\begin{description}
\item[RQ1] Virtually every group of authors defines their own control-flow
and authorization specification models. It is common to limit control-flow
support to either linear workflows (especially in the initial works) or
partial orders, although most recent works expand this to support also
conditional executions and even loops. Likewise, authorization constraints
are usually limited to Separation of Duties in its many forms. There are
attempts to classify authorization constraints, but there is still no common
framework to specify them.

\item[RQ2] The main related problem is workflow resiliency and there is
already a considerable amount of research on this topic. Besides that, all
the problems described in Section~\ref{sec:other} either have been identified
from the study of the WSP or can be solved by reusing WSP techniques.

\item[RQ3] It is clear that there is a trend of supporting more complex
control-flow and constraints. Current works already consider realistic
control-flow, but there is a need for more complex constraints (e.g.,
temporal, instance-spanning) and authorization policies. Developing a common
understanding and classification of authorization constraints in
security-sensitive workflows remains a challenge.
\end{description}

\section*{Acknowledgement}

This work has been partly supported by the EU under grant 317387 SECENTIS
(FP7-PEOPLE-2012-ITN).

\bibliographystyle{elsarticle-num}
\bibliography{biblio}

\end{document}